\relax
\documentclass[letterpaper]{article} 

\usepackage{aaai20}  
\usepackage{times}  
\usepackage{helvet} 
\usepackage{courier}  
\usepackage[hyphens]{url}  
\usepackage{graphicx} 
\usepackage{graphicx}  
\usepackage{bm}
\usepackage{soul}
\usepackage{url}
\usepackage[utf8]{inputenc}
\usepackage[small]{caption}
\usepackage{graphicx}
\usepackage{epstopdf}
\usepackage{amsmath}
\usepackage{booktabs}
\usepackage{algorithm,algpseudocode}
\usepackage{amsfonts,amssymb}
\usepackage{caption}
\usepackage{subfig}
\usepackage{subfloat}
\usepackage{comment}
\usepackage{amsmath}
\usepackage{multirow}

\title{Solving Cold Start Problem in Recommendation \\with Attribute Graph Neural Networks}
\author{\Large \textbf{Tieyun Qian\textsuperscript{\rm 1}, Yile Liang\textsuperscript{\rm 1}, Qing Li\textsuperscript{\rm 2}}\\ 
\textsuperscript{\rm 1}School of Computer Science, Wuhan University\\
\textsuperscript{\rm 2}Department of Computing, The Hong Kong Polytechnic University\\
\textsuperscript{\rm 1}\{qty,liangyile\}@whu.edu.cn, \textsuperscript{\rm 2}csqli@comp.polyu.edu.hk 
}

\begin{document}

\maketitle

\begin{abstract}
Matrix completion is a classic problem underlying recommender systems. It is traditionally tackled with matrix factorization. Recently, deep learning based methods, especially graph neural networks, have made impressive progress on this problem. Despite their effectiveness, existing methods focus on modeling the user-item interaction graph. The inherent drawback of such methods is that their performance is bound to the density of the interactions, which is however usually of high sparsity. More importantly, for a cold start user/item that does not have any interactions, such methods are unable to learn the preference embedding of the user/item  since there is no link to this user/item in the graph.

In this work, we develop a novel framework \emph{Attribute Graph Neural Networks} (AGNN) by exploiting the attribute graph rather than the commonly used interaction graph. This leads to the  capability of learning embeddings for cold start users/items. Our AGNN can produce the preference embedding for a cold user/item by learning on the distribution of attributes with an extended variational auto-encoder structure.  Moreover, we propose a new graph neural network variant, i.e., gated-GNN, to effectively aggregate various attributes of different modalities in a neighborhood. Empirical results on two real-world datasets demonstrate that our model yields significant improvements for cold start recommendations and outperforms or matches state-of-the-arts performance in the warm start scenario.
\end{abstract}

\section{Introduction}\label{sec:intro}
Matrix completion  is a well-known recommendation task aiming at predicting a user's ratings for those items which are  not rated yet by the user. Collaborative filtering (CF) ~\cite{CF_cacm99} has been successfully used to build recommender systems in various domains.
Matrix factorization (MF) ~\cite{MFT_com09} is one of the most prevalent method in CF due to its  high predicting performance and scalability. Given a $M$$\times$$N$ user-item rating matrix, MF first performs a low rank approximation to learn the user's and item's latent representation, also known as \emph{preference embedding} of a user or an item, and then uses a score function over the learnt preference embeddings to generate ratings for the missing entries in the matrix.
Sparsity and its extreme case of \emph{cold start, where  a user/item that does not have any interactions}, are the severe problem in recommender systems.
The performance of MF methods will drop quickly in the sparsity or the cold start settings. Conventional CF approaches to this issue are to generate \emph{feature embedding} using side information ~\cite{SocialRec_wsdm11,Rendle2011Factorization,FISM_kdd13,TrustSVD_aaai15,Zheng_DeepCoNN_wsdm2017,Chen_NARRE_www2018}. Such methods often introduce additional objective terms which make the learning and inference process very complicated.

Recent advances in deep learning, especially graph neural networks (GNNs), shed new light on this classic recommendation problem. The main advantage of GNN is that it can represent information from its neighborhood with arbitrary depth ~\cite{GCMC_kdd18,PinSage_kdd18,NGCF_sigir19,DiffNet_sigir19,DANSER_www19}.
GNN allows learning high-quality user and item representations, and consequently achieves the state-of-the-art performance.
However, almost all existing GNN based methods are built upon the user-item bipartite graph, where the node denotes a user or an item, and the edge is the interaction between the user and the item. Such methods cannot be used for cold start recommendations.


Indeed, little attention has been paid on using deep network architectures to address cold start issues. We are aware of several deep learning methods towards this problem, i.e.,  DropoutNet~\cite{DropoutNet_NIPS2017}, STAR-GCN~\cite{StarGCN_ijcai19}, and HERS~\cite{HERS_aaai19}.
Despite their effectiveness, both DropoutNet and STAR-GCN have an inherent limitation, i.e., their performance is bound to the number of interactions. The reason is that STAR-GCN relies on the interaction graph. It requires an ask-to-rate technique which might be not applicable to the real-world cold start scenario. Meanwhile, the objective of DropoutNet is  to reconstruct the rating of the user-item pair. That is to say, the training of the DropoutNet model is still dependent on the existing interactions.
Moreover, though HERS utilizes user-user and item-item relations to address cold start problem by referring to the influential nodes in contexts, the drawback is that it might recommend the popular item to the new user, or vice versa, as it represents  cold start nodes by neighbor aggregation without considering the new nodes' own attributes.

In order to address the above limitations, we propose a novel framework \emph{Attribute Graph Neural Networks} (AGNN) by exploiting the attribute graph instead of the widely used user-item graph.
Unlike the ratings, the attributes are available even for cold start  users/items. For example, when a merchant starts to sale its products online, it is necessary to provide the product attributes such as the category, description, and image. Similarly, many web-sites ask users to fill their profile information like gender and location at the time of registration.

While being ready to exploit the attribute information for cold start recommendations, there are two key challenges that hinder its potential. One is how to transform the attribute representation into the preference representation. The other is how to effectively aggregate attributes of different modalities, e.g., textual description and image, of the nodes in a neighborhood.
In this work, we first exploit \emph{an extended variational auto-encoder (eVAE) structure} to directly learn the preference embedding from the attribute distribution, with the perception that users' or items' preference can be inferred from their attributes. For example, a female user may prefer the romantic movie to the horror one.  We further design \emph{a gated-GNN structure} to  aggregate the complicated  node embeddings in the same neighborhood, which enables a leap in model capacity since it can assign different importance to each dimension of the node embeddings.

We conduct extensive experiments on two real-world datasets. Results demonstrate that our proposed AGNN model  yields significant improvements over the state-of-the-art baselines for cold start recommendations, and it also outperforms or matches the performance of these baselines in the warm start scenario.

\section{Related Work}
\textbf{Collaborative filtering (CF) methods }
CF is commonly used to leverage the user-item interaction data for recommendation. It mainly consists of neighbor-based methods~\cite{Sarwar2001Item,Koren_NeighborCF_kdd08} and matrix factorization (MF) methods~\cite{PMF_nips08,MFT_com09}. Recently, the CF approaches are extended with deep learning techniques~\cite{CDL_kdd15,NNMF_arxiv15,DropoutNet_NIPS2017,NFM_sigir17}.


\textbf{Graph neural  network (GNN) based methods }
The first GNN architecture employed for recommendation is GCN. RMGCNN ~\cite{RMGCNN_NIPS17} adopted GCN framework to aggregate information from user-user and item-item graphs. GCMC ~\cite{GCMC_kdd18} applied the graph convolutions on the user-item rating graph. PinSage ~\cite{PinSage_kdd18} combined efficient random walks and graph convolutions to generate embeddings of nodes. STAR-GCN ~\cite{StarGCN_ijcai19} designed a stacked and reconstructed GCN to improve the prediction performance.

The GNN architecture is mainly used  for the recursive diffusion in social recommendation~\cite{GraphRec_www19,DiffNet_sigir19}, and NGCF~\cite{NGCF_sigir19} encoded the high-order connectivity by performing embedding propagation. Finally, DANSER \cite{DANSER_www19} was the first to deploy the GAT for collaboratively modeling social effects for recommendation tasks.

Our proposed model differs from the previous GNN based methods in two issues. Firstly, our model employs the attribute graph instead of user-item interaction graphs. Almost all existing methods are based on the interaction graphs. The only exception DiffNet~\cite{DiffNet_sigir19} is based on users' social graph rather than the attribute graph in our model.
Secondly, we design a gated GNN structure to differentiate the importance of each dimension of node embeddings. Among the aforementioned methods, only DANSER took the importance of different nodes in the neighborhood into account. However, its model is at the node level. Due to the coarse granularity, it is hard for DANSER to fully leverages the power of GNN architectures.

\textbf{Dealing with Cold Start Issues }
Sparsity and cold start are prevalent in recommender systems. A promising approach is to leverage side information. Conventional methods mainly exploited side information as regularization in MF objective function ~\cite{SocialRec_wsdm11,TrustSVD_aaai15}. Recent studies focused on developing various types of neural networks to incorporate such information ~\cite{Zheng_DeepCoNN_wsdm2017,Chen_NARRE_www2018,DANSER_www19,RelationalCF_sigir19,RCDFM_aaai19}.

Our proposed AGNN has a similar neighbor aggregating architecture  with several deep learning based methods ~\cite{RMGCNN_NIPS17,DANSER_www19,StarGCN_ijcai19,HERS_aaai19} in addressing cold start issues. The key difference is in the eVAE structure which is uniquely used in AGNN to generate preference embedding from attribute embedding of different modality. To our knowledge, this is the first time that VAE~\cite{VAE_ICLR14} is used for this purpose, while previous researches in recommendation adopted VAE to reconstruct the latent representations with the same  modality~\cite{CVAE_kdd17,MultVAE_www18,SVAE_wsdm19}.

\section{Proposed Model}

\subsection{Problem Definition}
Let $U = \{u_1, u_2, ..., u_M\}$ and $V = \{v_1, v_2, ..., v_N\}$ be the set of
users and items, and $M$ and $N$ is their corresponding cardinality, respectively. In addition, each user or item is associated with a set of attributes from different fields. Each attribute value has a separated encoding, and all attributes are concatenated into a multi-hot attribute encoding $\mathbf{a} \in \mathbb{R}^K$. Below is an example of user attribute encoding $\mathbf{a}_u$.

\begin{equation} \nonumber
\small
\mathbf{a}_u =\underbrace{[0, 1]}_{\textit{gender}} \ \underbrace{[1, 0, 0,... , 0]}_{\textit{age}} \ \underbrace{[0, 1, 0,... , 0]}_{\textit{occupation}},
\end{equation}
Let $\mathbf{R} \in \mathbb{R}^{M \times N}$ be the user-item interaction matrix, which consists of real-valued ratings for explicit interactions, or binary entries for implicit feedbacks such as click or not. In this paper, we tackle the recommendation task with explicit interactions, where each  $r_{ij} \in \mathbf{R}$ is either a rating score denoting $u_i$ gives a rating to $v_j$, or 0 denoting the unknown ratings of items that the users have not interacted yet. The goal is to predict these unknown ratings.
In particular, we are interested in \emph{the cold start rating prediction problem}, i.e., there is no preference information available for the cold start users (items) except their attribute information.

\begin{figure*}[t]
	\vspace{-2mm}
	\centering
	\includegraphics[width=0.85\textwidth]{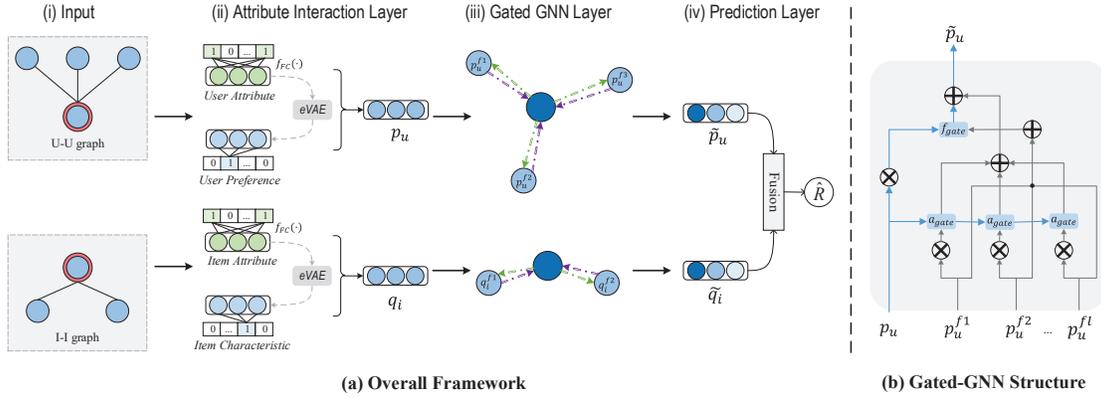} 
	\vspace{-2mm}
	\caption{On the left is the framework of our model; on the right is an illustration the gated-GNN structure.}
\vspace{-2mm}
	\label{fig1}
\vspace{-1mm}
\end{figure*}

\subsection{Model Overview}
The architecture of the proposed AGNN model  is shown in Figure \ref{fig1}. It consists of an input layer, an interaction layer, a gated-GNN layer, and a prediction layer.

We first design  an input layer to build the user  (item) attribute graph $\mathcal{A}_u$ ($\mathcal{A}_i$).
We then present an interaction layer to integrate one node's different information into a unified embedding.  
We also  develop a key eVAE component to generate the preference embedding for cold start nodes in this layer.
Next, we propose a gated-GNN layer to aggregate the complicated node embeddings in a neighborhood in the attribute graph.
Finally, we add a prediction layer to let the aggregated representations of user and item interact with each other to calculate the rating score.

\subsection{Model Architecture}
\paragraph{Input layer } Our model differs from exiting ones in that it is upon the homogeneous attribute graph rather the bipartite user-item graph.  This enables our model to free from sparse interactions and to deal with the cold start problem.

We construct the attribute graph  using  attribute information in this layer.
The quality of attribute graph plays an important role in our task. However, a detailed analysis of how to construct good graphs is beyond the scope of this paper. Hence we resort to a natural proximity-based way to construct the attribute graph. We first define two types of proximities, i.e., \textit{preference proximity} and \textit{attribute proximity}.


(1) The preference proximity measures the historical preference similarity between two nodes. If two users have similar rating record list (or two items have similar rated record list), they will have a high preference proximity. Note we cannot calculate preference proximity for the cold start nodes as they do not have the historical ratings.

(2) The attribute proximity measures the  similarity between the attributes of two nodes. If two users have similar user profiles, e.g., gender, occupation (or two items have similar properties, e.g., category), they will have a high attribute proximity.

Both types of proximity can be measured by cosine distance. It is calculated as:
\begin{equation}\label{equ:cossim}
\small
\vspace{-2mm}
proximity(\mathbf{w}, \mathbf{v}) = 1 - \frac{\mathbf{w} \cdot \mathbf{v}^T} { \Vert\mathbf{w}\Vert \Vert\mathbf{v}\Vert}, 
\vspace{-0mm}
\end{equation}
where $\mathbf{w}$ and $\mathbf{v}$ are two nodes' preference representations or their multi-hot attribute encodings.
Two types of proximity are summed after the min-max normalization to get an overall proximity.

After calculating the overall proximity between two nodes, it becomes a natural choice to build a \emph{k}-NN graph as adopted in \cite{RMGCNN_NIPS17}. Such a method will keep a fixed number  of neighbors once the graph is constructed. It may work well when the graph is constructed on the single type of node attribute like a social graph. However, since our similarity is defined on multiple types of attributes,  it is necessary to maintain a diversity of neighborhood to some extent. The rationale is that  we wish the age is the dominant factor in determining the neighborhood in some cases while the occupation holds the lead in other cases.
To this end, we propose \emph{a dynamic graph construction strategy}. To be specific, for a node $u$, we add all the nodes which have a top $p\%$ proximity with node $u$ to the candidate pool $N_u^C$. During each round of the training process, the neighbors of node $u$ are sampled according to the proximity from the candidate pool.

\vspace{-3mm}
\paragraph{Attribute Interaction Layer }
In the constructed attribute graph $\mathcal{A}_u$ and $\mathcal{A}_i$, each  nodes has an attached multi-hot attribute encoding and a unique one-hot representation denoting its identity.
Due to the huge number of users and items in the web-scale recommender systems, the dimensionality of nodes' one-hot representation is extremely high. Moreover, the multi-hot attribute representation  simply combines multiple types of attributes into one long vector without considering their interactive relations.

The goal of interaction layer is to reduce the dimensionality for  one-hot identity representation and learn the high-order attribute interactions for multi-hot attribute representation.
To this end, we first set up a lookup table to transform a node's one-hot representation into the low-dimensional dense vector. The lookup layers correspond to two parameter matrices $\mathbf{M} \in \mathbb{R}^{M \times D}$ and $\mathbf{N} \in \mathbb{R}^{N \times D}$. Each entry $\mathbf{m}_u$ $\in$ $\mathbb{R}^D$ and $\mathbf{n}_i$ $\in$ $\mathbb{R}^D$ encodes the user $u$'s preference and the item $i$'s property, respectively.
Note that $\mathbf{m}_u$ and $\mathbf{n}_i$ for cold start nodes are meaningless, since no interaction is observed to train their preference embedding. We will discuss the solution to this problem later.

Inspired by ~\cite{NFM_sigir17}, we capture the high-order attribute interactions with a Bi-Interactive pooling operation, in addition to the linear combination operation. To be specific, let $\mathbf{v}_i$ and $\mathbf{v}_j$ be the embedding vector for the $i$-th and $j$-th type of attribute in the multi-hot attribute encoding $\mathbf{a} \in \mathbb{R}^K$, respectively,  the Bi-Interactive and linear combination operation are defined as:
\begin{equation}\label{equ:bi-interaction}
\vspace{-1mm}
\small
f_{BI}(\mathbf{a}) = \sum_{i=1}^{K}\sum_{j=i+1}^{K} a_i\mathbf{v}_i \odot a_j\mathbf{v}_j, \quad f_{L}(\mathbf{a}) = \sum_{i=1}^{K} a_i\mathbf{v}_i,
\vspace{-1mm}
\end{equation}
where $\odot$ denotes the element-wise product.

Finally, a fully connected layer is added on both the second-order interaction and linear combination to learn the high-order feature interactions:
\begin{equation}\label{equ:FC}
\small
f_{FC}(\mathbf{a}) =  LeakyReLU (\mathbf{W}_{fc}^{(1)} f_{BI}(\mathbf{a}) + \mathbf{W}_{fc}^{(0)} f_{L}(\mathbf{a}) + \mathbf{b}_{fc}),
\end{equation}
where $\mathbf{W}_{fc}, \mathbf{b}_{fc},  LeakyReLU$ are weight matrix, bias vector, and activation function, respectively.
We can then get the attribute embedding $\mathbf{x}_u$ and $\mathbf{y}_i$ for a user $u$ and for an item $i$  by feeding their respective attribute encoding $\mathbf{a}_u$ and $\mathbf{a}_i$ into the $f_{FC}$ function, i.e.,
\begin{equation}\label{equ:final_xuyi}
\small
\mathbf{x}_u = f_{FC}(\mathbf{a}_u),\quad \mathbf{y}_i = f_{FC}(\mathbf{a}_i)
\end{equation}

Next, we fuse the preference embedding and attribute embedding into the node embedding such that each node  contains both historical preferences and its own attributes.
\vspace{-2mm}
\begin{equation}\label{equ:fuse_attr1}
\small
\mathbf{p}_u = \mathbf{W}_u [\mathbf{m}_u;\ \mathbf{x}_u] + \mathbf{b}_u,\quad
\mathbf{q}_i = \mathbf{W}_i [\mathbf{n}_i;\ \mathbf{y}_i] + \mathbf{b}_i,
\end{equation}
where $[;]$ denotes vector concatenation operation, $\mathbf{W}_{u(i)}$, $\mathbf{b}_{u(i)}$ are weight matrix and bias vector.
For cold start nodes without any interactions, we will generate preference embeddings for them in this layer. We will detail our solution to this cold start problem in a separate subsection later.

\vspace{-3mm}
\paragraph{Gated GNN Layer }

Intuitively, different neighbors have different relations to a node. Furthermore, one neighbor usually has multiple attributes. For example, in a social network, a user's neighborhood may consist of classmates, family members, colleagues, and so on, and each neighbor may have several attributes such as age, gender, and occupation. Since all these attributes (along with the preferences) are now encoded in the node's embedding, it is necessary to pay different attentions to different dimensions of the neighbor node's embedding. However, existing GCN ~\cite{GCN_iclr17} or GAT ~\cite{GAN_iclr18} structures cannot do this because they are at the coarse granularity. GCN treats all neighbors equally and GAT differentiates the importance of neighbors at the node level. To solve this problem, we design a gated-GNN structure to aggregate the fine-grained neighbor information.

Our proposed gated-GNN structure is shown in Figure \ref{fig1} (b). It contains an \textit{aggregate gate} (denoted as $\mathbf{a}_{gate}$)  and a \textit{filter gate} (denoted as $\mathbf{f}_{gate}$). In order to better capture the homophily phenomenon in networks, the $\mathbf{a}_{gate}$ controls what information should be aggregated from neighbors to the target node, while $\mathbf{f}_{gate}$ controls what information in the target node should be filtered out if it is not consistent with that in the neighbors. These two gates work as follows.

Given a user node $u$, its node embedding $\mathbf{p}_u$, its neighbor set $N_u$, and the node embedding $\mathbf{p}_u^{f_i}$ for the $i$-th neighbor $f_i$ in $N_u$, we first apply $\mathbf{a}_{gate}$ to the neighbors to obtain the aggregated representation $\mathbf{p}_{u}^{u\leftarrow N_u}$ by selectively passing the neighbor embeddings to the target node $u$.
\begin{equation}\label{equ:aggregate_gate}
\small
\mathbf{a}_{gate}^{f_i} = \sigma (\mathbf{W}_a [\mathbf{p}_u; \ \mathbf{p}_u^{f_i}] + \mathbf{b}_a),
\end{equation}
\vspace{-3mm}
\begin{equation}\label{equ:nei_aggregate}
\small
\mathbf{p}_{u}^{u\leftarrow N_u} =  \frac{1}{|N_u|} \sum_{i=1}^{|N_u|} ( \mathbf{p}_u^{f_i} \odot \mathbf{a}_{gate}^{f_i}),
\end{equation}
where $\mathbf{W}_{a}, \mathbf{b}_{a},  \sigma$ are weight matrix, bias vector, and the sigmoid activation function.

We then apply the filter gate $\mathbf{f}_{gate}$ to the target node $u$ to filter out its  information that is inconsistent with the averaged  representations of the neighbors. More formally,
\vspace{-1mm}
\begin{equation}\label{equ:filter_gate}
\small
\mathbf{f}_{gate} = \sigma (\mathbf{W}_f [\mathbf{p}_u; \ \frac{1}{|N_u|} \sum_{i=1}^{|N_u|} \mathbf{p}_u^{f_i}] + \mathbf{b}_f),
\end{equation}
\vspace{-3mm}
\begin{equation}\label{equ:target_filter}
\small
\mathbf{p}_{u}^{-} =  \mathbf{p}_u \odot (\mathbf{1} - f_{gate}),
\end{equation}
where $\mathbf{p}_{u}^{-}$ is the node $u$'s remaining representation after the filtering operation.

Combining the aggregated representation $\mathbf{p}_{u}^{u\leftarrow N_u}$ and the  remaining representation $\mathbf{p}_{u}^{-}$ together, we can get the user node $u$'s final embedding $\mathbf{\widetilde{p}}_{u}$ as follows:
\begin{equation}\label{equ:gnn_final}
\small
\mathbf{\widetilde{p}}_{u} =  LeakyReLU(\mathbf{p}_{u}^{-} + \mathbf{p}_{u}^{u\leftarrow N_u})
\end{equation}

The item $i$'s final embedding, denoted as $\mathbf{\widetilde{q_i}}$,  can be obtained from the item attribute graph in a similar way.

\vspace{-3mm}
\paragraph{Prediction Layer }
Given a user $u$'s final representation $\mathbf{\widetilde{p}}_{u}$ and an item $i$'s final representation $\mathbf{\widetilde{q}}_{i}$  after the gated-GNN layer, we model the predicted rating of the user $u$ to the item $i$ as:
\begin{equation}\label{equ:predict}
\small
\hat{R}_{u,i} = MLP([\mathbf{\widetilde{p}}_u;\ \mathbf{\widetilde{q}}_i]) + \mathbf{\widetilde{p}}_u \mathbf{\widetilde{q}}_i^T + b_u + b_i + \mu,
\end{equation}
where the MLP function is the multi-layer perception implemented with one hidden layer, and $b_u$, $b_i$, and $\mu$ denotes user bias, item bias, and global bias, respectively. In Eq.~\ref{equ:predict}, the second term is  inner product interaction function ~\cite{MFT_com09}, and we add the first term  to capture the complicated nonlinear interaction between the user and the item.

\vspace{-3mm}
\paragraph{Solution to Cold Start Problem }
The cold start problem is caused by the lack of historical interactions for cold start nodes. We view this as a missing preference problem. Unlike the methods in ~\cite{DropoutNet_NIPS2017,StarGCN_ijcai19} which reconstruct the same node embedding  with the dropout or mask technique, we aim to  reconstruct the node's missing preference embedding from its attribute embedding.

Basically, one specific type of users might be interested in the similar items, and vice versa. For example, animation is the mainstream entertainment among teenage children (the users have the similar age attribute). This indicates that the attribute and preference embeddings are not only close to each other in the latent space but also have the similar distribution.
Hence we tackle the missing preference problem by employing the variational auto-encoder structure to reconstruct the preference from the attribute distribution.

\begin{figure}[t]
\vspace{-0mm}
	\centering
	\includegraphics[width=0.76\columnwidth]{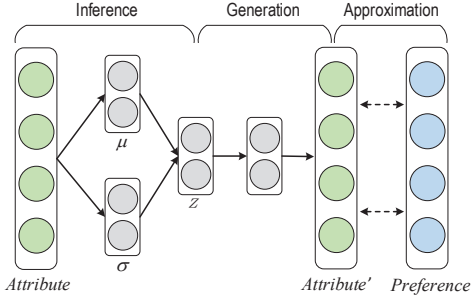} 
\vspace{-0mm}
	\caption{The eVAE structure to generate preference embedding from attribute distribution.}
\vspace{-3mm}
	\label{fig2}
\vspace{-3mm}
\end{figure}

Our proposed extended VAE (eVAE) structure is shown in Figure \ref{fig2}, which contains three parts: inference, generation, and approximation. The first two parts are the standard VAE and the third one is our extension. Take the cold start user node $u$ as an example.
In the generation part, $u$ is given a latent variables $\mathbf{z}_u$. The reconstructed embedding $\mathbf{x}'_u$ is generated from its latent variables $\mathbf{z}_u$ through generation network as MLP parameterized by $\theta$:
\begin{equation}\label{equ:vae_generate}
\small
\mathbf{x}'_u \sim p_\theta(\mathbf{x}'_u | \mathbf{z}_u).
\end{equation}

In the inference part, variational inference approximates the true intractable posterior of the latent variable $\mathbf{z}_u$ by introducing an inference network parameterized by $\phi$:
\begin{equation}\label{equ:vae_inference}
\small
q_\phi(\mathbf{z}_u) = \mathcal{N}(\bm{\mu}_u,\ diag(\bm{\sigma}_u^2)),
\end{equation}
The objective of variational inference is to optimize the free variational parameters so that the  KL-divergence $KL(q(\mathbf{z}_u )\Vert p(\mathbf{z}_u|\mathbf{x}_u))$ is minimized.
With the reparameterization trick \cite{VAE_ICLR14}, we sample $\bm{\epsilon} \sim \mathcal{N}(0, \mathbf{I})$ and reparameterize $\mathbf{z}_u = \mu_\phi(\mathbf{x}_u) + \bm{\epsilon} \odot \sigma_\phi(\mathbf{x}_u)$. In this case, the gradient towards $\phi$ can be back-propagated through the sampled $\mathbf{z}_u$.

In the approximation part, we constrain the reconstructed embedding $\mathbf{x}'_u$  to be close to the preference embedding $\mathbf{m}_u$.
This is practical because the system should have collected a certain amount of interactions in reality.  During the training phase, the nodes with historical ratings actually have the preference embeddings. Such information can be explored to improve the VAE. Hence we require the reconstructed embedding $\mathbf{x}'_u$ to be similar with both the preference embedding (by the constraint) and the original attribute distribution (by the standard VAE).
To summarize, the reconstruction  loss function in our proposed eVAE is defined as follows.
\begin{equation}\label{equ:vae_loss}
\small
\begin{split}
\mathcal{L}_{recon} =  &-KL(q_\phi(\mathbf{z}_u |\mathbf{x}_u)\Vert p(\mathbf{z}_u)) \\
&+ \mathbb{E}_{q_\phi(\mathbf{z}_u |\mathbf{x}_u)}[\log p_\theta(\mathbf{x}'_u |\mathbf{z}_u)] + \Vert \mathbf{x}'_u - \mathbf{m}_u \Vert_2,
\end{split}
\end{equation}
where the first two terms are same as those in standard VAE, and the last one is our extension for the approximation part. The cold start item $i$'s preference embedding $\mathbf{n}_i$ can be generated similarly from its attribute embedding $\mathbf{y}'_i$, and thus we have $\mathbf{m}_u \sim \mathbf{x}'_u$ and $\mathbf{n}_i \sim \mathbf{y}'_i$.



\vspace{-3mm}
\paragraph{Loss } The overall loss function for training is defined as:
\begin{equation}\label{equ:loss}
\vspace{-1mm}
\small
\mathcal{L} = \mathcal{L}_{pred} + \mathcal{L}_{recon},
\vspace{-1mm}
\end{equation}
where $\mathcal{L}_{pred}$ is the task-specific rating prediction loss, and $\mathcal{L}_{recon}$ is the reconstruction loss defined in Eq. ~\eqref{equ:vae_loss}.

For the rating prediction loss, we employ the square loss as the objective function:
\begin{equation}\label{equ:pred_loss}
\small
\mathcal{L}_{pred} = \sum\nolimits_{u,i \in \mathcal{T}}(\hat{R}_{u,i} - R_{u,i})^2,
\end{equation}
where $\mathcal{T}$ denotes the set of instances for training, i.e., $\mathcal{T}$ = \{($u$, $i$, $r_{u,i}$, $\mathbf{a}_u$, $\mathbf{a}_i$)\}, $R_{u,i}$ is ground truth rating in the training set $\mathcal{T}$, and $\hat{R}_{u,i}$ is the predicted rating.

\section{Experiments}

\subsection{Experimental Setup}
\paragraph{Datasets}
We use two publicly available datasets to evaluate our model. We employ the ML-100K version of \textbf{MovieLens\footnote{https://grouplens.org/datasets/movielens/}} dataset. We extend it by crawling stars, directors, writers and countries from IMDb\footnote{https://www.imdb.com} according to the movie title and release year. We take categories, stars, directors, writers, and countries as movie features, and gender, age, and occupations as user features.  We pre-process \textbf{Yelp\footnote{https://www.yelp.com/dataset\_challenge}} dataset by removing nodes with less than 20 ratings. We take categories, located states, and  located cities as item features, and use social links as user-user graph and also as attributes for users due to the lack of profile information on Yelp.  The statistics of two datasets are shown in Table \ref{tab:dataset}.
\begin{table}[]
	\caption{Statistics of the evaluation datasets.}
	\label{tab:dataset}
\vspace{-2mm}
	\centering
	\small
	\renewcommand\arraystretch{1.1}
	\begin{tabular}{c|c|c|c|c}
		\hline	
		Datasets & \#User & \#Item & \#Rating & \#Sparsity \\ \hline
		MovieLens & 943 & 1682 & 100000 & 93.70\% \\ \hline
		Yelp & 23549 & 17139 & 941742 & 99.77\% \\ \hline
	\end{tabular}
\vspace{-1mm}
\end{table}
\vspace{-1mm}

\begin{table*}[]
	\caption{Performance comparison on two datasets. The best performance among all is in bold while the best one among baselines is marked with an underline.  The last row indicates the percentage of improvements gained by the proposed method compared with the best baseline.}
	\label{tab:performance}
	\centering
	\small
	\def\arraystretch{1.2}
	\begin{tabular}{l|c|c|c|c|c|c|c|c|c|c|c|c}
		\hline
		\multirow{3}{*}{Method} & \multicolumn{6}{c|}{MovieLens} & \multicolumn{6}{c}{Yelp} \\ \cline{2-13}
		& \multicolumn{2}{c|}{item cold start} & \multicolumn{2}{c|}{user cold start} & \multicolumn{2}{c|}{warm start} & \multicolumn{2}{c|}{item cold start} & \multicolumn{2}{c|}{user cold start} & \multicolumn{2}{c}{warm start} \\ \cline{2-13}
		& RMSE & MAE & RMSE & MAE & RMSE & MAE & RMSE & MAE & RMSE & MAE & RMSE & MAE \\ \hline
		NFM & 1.0416 & 0.8525 & 1.0399 & 0.8404 & 0.9533 & 0.7565 & 1.1231 & 0.9077 & 1.1045 & 0.8832 & 1.0620 & 0.8372 \\ \hline
		DropoutNet & 1.0844 & 0.8722 & 1.0654 & 0.8571 & 0.9428 & 0.7399 & 1.1891 & 0.9628 & 1.1724 & 1.9624 & 1.1524 & 0.9254 \\ \hline
		DiffNet & 1.0418 & 0.8476 & \underline{1.0379} & \underline{0.8380} & 0.9221 & 0.7250 & \underline{1.1072} & \underline{0.9012} & 1.1267 & 0.9144 & 1.0444 & 0.8241 \\ \hline
		DANSER & 1.1190 & 0.9414 & 1.0490 & 0.8542 & 0.9823 & 0.7830 & 1.1302 & 0.9095 & \underline{1.0927} & \underline{0.8818} & 1.0525 & 0.8319 \\ \hline
		sRMGCNN & 1.1532 & 0.9434 & 1.0479 & 0.8411 & 0.9376 & 0.7458 & -- & -- & -- & -- & -- & -- \\ \hline
		GC-MC & 1.0392 & 0.8470 & 1.0444 & 0.8647 & 0.9106 & 0.7150 & 1.1229 & 0.9111 & 1.1020 & 0.9235 & 1.0254 & 0.8205 \\ \hline
        STAR-GCN & \underline{1.0376} & \underline{0.8440} & 1.0428 & 0.8596 & \underline{\textbf{0.9049}} & \underline{\textbf{0.7116}} & 1.1173 & 0.9088 & 1.0988 & 0.9162 & \underline{1.0232} & \underline{0.8201} \\ \hline
		AGNN & \textbf{1.0187} & \textbf{0.8171} & \textbf{1.0208} & \textbf{0.8198} & 0.9078 & 0.7138 & \textbf{1.0749} & \textbf{0.8715} & \textbf{1.0657} & \textbf{0.8586} & \textbf{1.0106} & \textbf{0.7945} \\ \hline
		Improvement & 1.82\% &	3.19\% & 	1.65\% & 	2.17\% & 	-0.32\% & 	-0.31\% & 	2.92\% & 	3.30\% & 	2.47\% & 	2.63\% & 	1.23\% &	3.12\%  \\ \hline
	\end{tabular}
\vspace{-2mm}
\end{table*}

\vspace{-0mm}
\paragraph{Baselines } We choose the following seven state-of-the-art methods as our baselines.
\begin{itemize}
 \item \textbf{NFM} \cite{NFM_sigir17} combines the linearity of FM and the non-linearity of NN into one framework.
 \item \textbf{DropoutNet} \cite{DropoutNet_NIPS2017} applies dropout technique to cold start problem.
 \item \textbf{sRMGCNN} \cite{RMGCNN_NIPS17} employs multi-graph convolutional neural network architecture for matrix completion.
 \item \textbf{GC-MC} \cite{GCMC_kdd18} adopts a GCN framework on user-item graph for matrix completion.
 \item \textbf{DiffNet} \cite{DiffNet_sigir19} includes a GCN-alike layer-wise diffusion procedure to model dynamic social diffusion in social recommendation.
 \item \textbf{DANSER} \cite{DANSER_www19} is a GAT-based method for social recommendation.
 \item \textbf{STAR-GCN} \cite{StarGCN_ijcai19} is a stacked GCN model and addresses cold start problem with mask  technique.
\end{itemize}

\vspace{-0mm}
\paragraph{Evaluation Metrics} We adopt the commonly used Rooted Mean Square Error (RMSE) and Mean Absolute Error (MAE) as the evaluation metrics.

\vspace{-0mm}
\paragraph{Settings}
We examine the  model performance in both the cold and warm start scenario. For cold start, we randomly choose 20\% items (or users) along with their interactions as test set, and the remaining interactions as training set. For warm start, we randomly choose 20\% user-item interactions as test set and the remaining 80\% as training set. The difference is that for cold start nodes, their interactions are totally removed from training.

The hyper-parameter settings in our AGNN are as follows: batch size = 128,  embedding dimension $D$ = 30,  initial learning rate = 0.0005, slop of LeakyReLU = 0.01, threshold $p$ in graph construction = 5. We use Adam ~\cite{KingmaB14} as optimizer to self-adapt the learning rate.

For the baselines, we strictly follow the same hyper-parameter settings if they are reported by authors.  The baselines designed for top-N recommendation are revised to optimize  RMSE scores. Please note that all baselines use the same attribute information as our model. We implement sRMGCNN wih its public source code, but it cannot scale to large dataset like Yelp. Besides, since DANSER is not designed for incorporating attributes, we take the attribute features to initialize its embedding for users and items. For DANSER and DiffNet, we remove the part for modeling social relationship on  MovieLens since there is no such information. Finally, we do not add newly rated edges to the cold start nodes in the testing phase of STAR-GCN, for a fair comparison with all other methods and also for simulating the real world cold start scenario.

\subsection{Comparison with Baselines}
The performance of AGNN and the baselines on both datasets are reported in Table \ref{tab:dataset}. We have the following important notes.

  (1) It is clear that our  AGNN outperforms all baselines in the cold start scenario. In particular, it achieves an improvement over the strongest baseline with a 1.82\% and 1.65\% RMSE score on MovieLens, and a 2.92\% and 2.47\% RMSE score on Yelp, for item and user cold start, respectively. The results verify the superiority of our proposed architecture by exploring attribute graph for cold start recommendation. Moreover, in the warm start scenario, our AGNN also yields the best results on Yelp and is the second best on MovieLens with a performance slightly inferior to STAR-GCN.

  (2) Among the baselines, sRMGCNN, GC-MC, and STAR-GCN utilize graph convolutional network on the user-item graph. STAR-GCN gets the overall best performance  because it integrates the content information into node embedding and also because it avoids the leakage issue when convoluting on user-item graph. The performance of GC-MC is limited as it incorporates content information after the convolution layer. sRMGCNN is the worst as it uses attributes to construct user-user or item-item graph without including them into the convolution operation. Moreover, it cannot handle large dataset like Yelp as its convolution is defined on Chebyshev expansion.

  (3) Two baselines DiffNet and DANSER utilize social graph for recommendation. DiffNet performs better in most cases because it combines user embedding with preference and attribute information. DANSER constructs item-item graph according to the number of co-purchased items. This results in its poor performance in item cold start. DropoutNet and NFM do not employ graph convolution operations. NFM performs well in many cases due to its ability to learn high-order feature interactions. DropoutNet is not good since it requires the content information to approximate the results of matrix factorization, and its performance  is dependent on the pre-trained preference embeddings.

In summary, while utilizing both attribute and structure information may improve the recommendation performance, the methods with shallow interactions between attribute and structure, e.g., sRMGCNN and GC-MC, are less effective than those with deep interactions, e.g., DiffNet, STAR-GCN, and AGNN. Furthermore, though STAR-GCN and DropoutNet adopt mask and dropout techniques for addressing cold start problem, they are built upon the user-item graph. The nature of the interaction graph makes them hard to achieve competitive  performance with our AGNN which exploits attribute graph with an eVAE structure.


\subsection{Ablation Study}

In order to verify the effectiveness of the key components in our model, we perform three types of ablation study and show their results in Table \ref{tab:ablation}. Due to the space limitation, we only present the RMSE scores and omit the MAE ones which are similar.
\begin{table}[]
	\vspace{-3mm}
	\caption{Results for ablation study in terms of RMSE.}
	\label{tab:ablation}
	\small
	\centering
	\def\arraystretch{1.1}
	\begin{tabular}{l|c|c|c|c}
		\hline
		\multirow{2}{*}{} & \multicolumn{2}{c|}{MovieLens} & \multicolumn{2}{c}{Yelp} \\ \cline{2-5}
		& cold item & cold user & cold item & cold user \\ \hline
		AGNN & \textbf{1.0187} & \textbf{1.0208} & \textbf{1.0749} & \textbf{1.0657} \\ \hline
		AGNN$_\text{knn}$ & 1.0298 & 1.0282 & 1.0805 & 1.0762 \\
		AGNN$_\text{cop}$ & 1.0717 & 1.0310 & 1.0788 & 1.0734 \\ \hline
		AGNN$_\text{GCN}$ & 1.0308 & 1.0280 & 1.0772 & 1.0766 \\
		AGNN$_\text{GAT}$ & 1.0262 & 1.0274 & 1.0768 & 1.0811 \\ \hline
		AGNN$_\text{mask}$ & 1.0230 & 1.0250 & 1.0847 & 1.0687 \\
        AGNN$_\text{drop}$ & 1.0256 & 1.0246 & 1.0885 & 1.0719 \\
		AGNN$_\text{-eVAE}$ & 1.0263 & 1.0253 & 1.0924 & 1.0724 \\ \hline
	\end{tabular}
\vspace{-3mm}
\end{table}

\textit{The first is to examine effects of graph construction method.}
	We compare AGNN with its two variants using different graph construction adapted from sRMGCNN and DANSER. Specifically, AGNN$_\text{knn}$ constructs user-user and item-item graph by choosing 10-nearest neighbors in the user and item attribute space only. AGNN$_\text{cop}$ constructs item-item  graph according to the number of co-purchased items. The user-user graph is constructed in a similar way if social links are not available.
	
	The performance of AGNN$_\text{cop}$ declines dramatically on Movielens since there is no neighbor for cold start nodes.  Its performance for cold start users does not change much on Yelp because social links already form the user-user graph. The superior performance of AGNN over AGNN$_\text{knn}$ demonstrates that both attribute and preference information are useful for graph construction. Moreover, AGNN benefits from its dynamic construction strategy as it allows to access diversified neighbors, and thus yields better performance than two variants with the fixed neighbors.
	
\textit{The second is to investigate effects of gated-GNN structure.}
    We compare AGNN with its two variants using different GNN structures adapted from GC-MC and DANSER. In particular, AGNN$_\text{GCN}$ employs an ordinary GCN by aggregating all neighbors' representations with a summation operation. AGNN$_\text{GAT}$ adopts an attention layer to learn the weight of each neighbor before aggregation.
	
	As can be seen, AGNN$_\text{GCN}$ is inferior to AGNN$_\text{GAT}$ in most of the cases, indicating that attention mechanism can improve the performance. However, they are both worse than AGNN. This verifies that differentiating the importance of each dimension of the node can further enhance the performance since it greatly enlarges the model capacity.
	
\textit{The third  is to investigate effects of eVAE structure.}
	We compare AGNN with its three variants, two adopting different strategies for cold start problem adapted from STAR-GCN and DropoutNet, and one by removing eVAE structure from our own AGNN. To be specific, AGNN$_\text{mask}$ randomly masks 20\% of the input nodes  and adds a decoder after gated-GNN layer to reconstruct the initial input node embedding. AGNN$_\text{drop}$ randomly sets 20\% preference embedding of the input nodes to 0. AGNN$_\text{-eVAE}$ is a simplified version by removing eVAE from AGNN.
	
	The performance of AGNN$_\text{-eVAE}$ is the worst since it does not contain component for dealing with cold start nodes after removing the eVAE structure. In addition, our complete AGNN with eVAE structure outperforms AGNN$_\text{mask}$ and AGNN$_\text{drop}$.  This demonstrates that the proposed eVAE structure is critical to our AGNN model,  and it is more effective than the mask and dropout techniques in addressing cold start problem.

\subsection{Performance Comparison w.r.t. Cold Start Ratio}
In the cold start scenario, a higher ratio of cold start nodes indicates that fewer user-item interactions can be utilized for collaborative filtering. This subsection compares our model with two strongest baselines, i.e., DiffNet and STAR-GCN, to  examine the impacts of the ratio of cold start nodes.

\begin{figure}[t]
	\vspace{-3mm}
	\centering
	\subfloat[item cold start\label{sfig:testa}]{%
		\includegraphics[width=0.49\columnwidth]{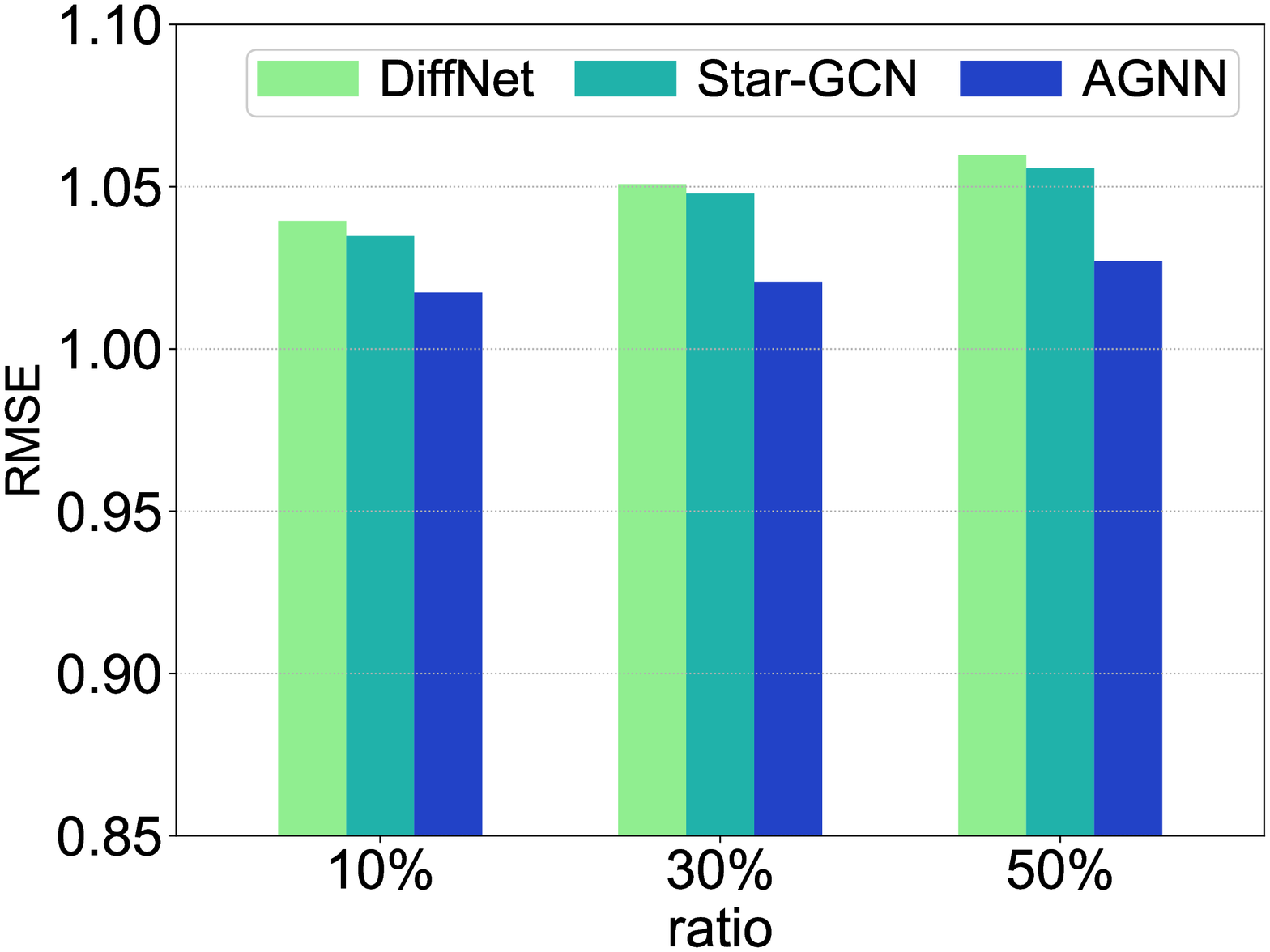}%
	}
	\subfloat[user cold start\label{sfig:testb}]{%
		\includegraphics[width=0.49\columnwidth]{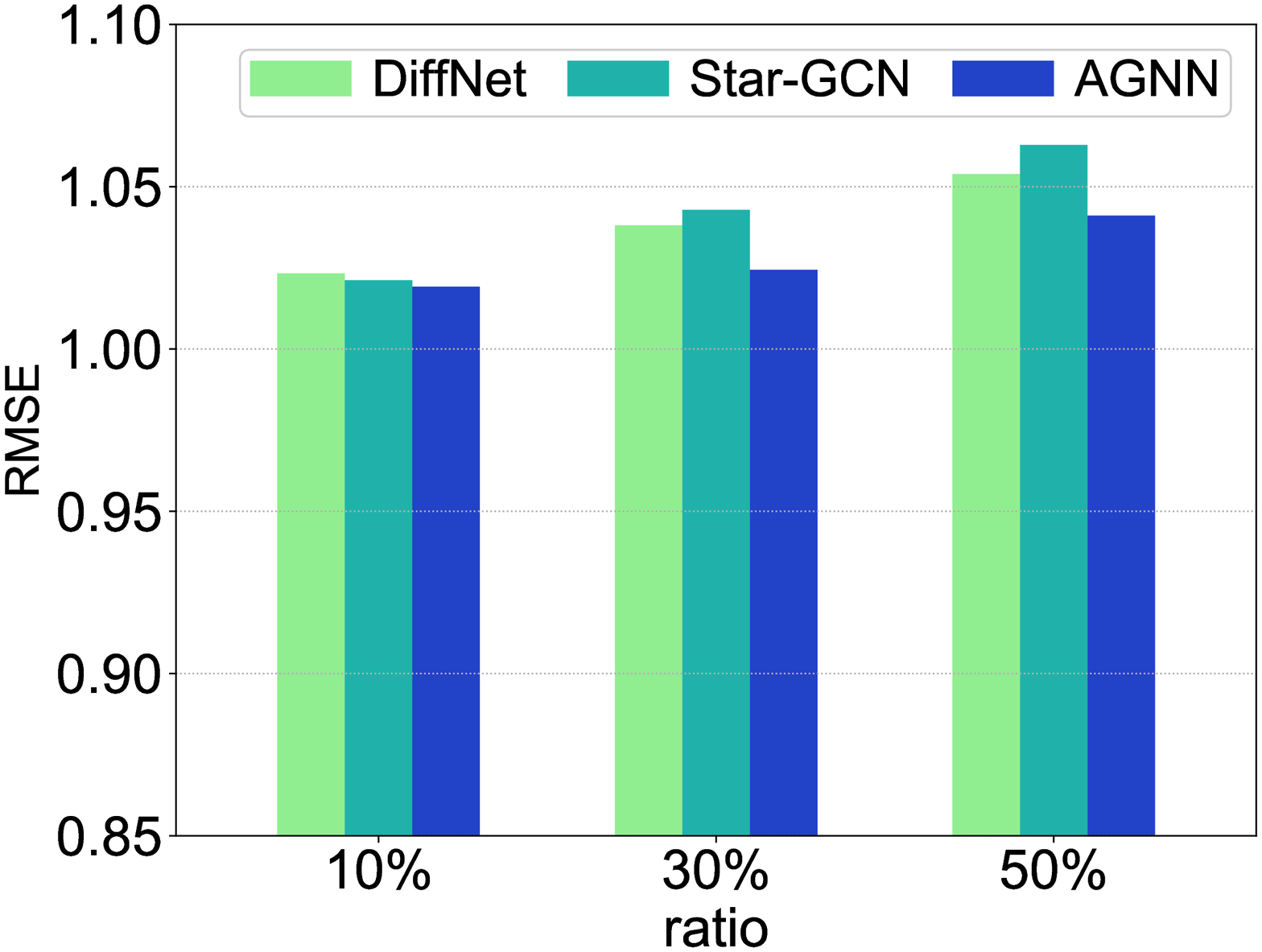}%
	}
	\vspace{-0mm}
	\caption{The performance comparison with varying percent of testing data in cold start scenario.}
	\label{fig:partition_ics}
\vspace{-2mm}
\end{figure}

We randomly choose 10\%, 30\% and 50\% nodes along with their interactions as test set, and the remaining interactions as training set. Due to the space limitation, we only report the RMSE results on MovieLens.  Figure \ref{fig:partition_ics} (a) and (b) show the results in item and user cold start scenario, respectively.  From the results, we have the following findings.

AGNN consistently outperforms DiffNet and STAR-GCN in different portions of cold start nodes.  This proves that the performance of our AGNN is stable among various cold start settings. 	
More importantly, when increasing the ratio of cold start nodes in the graph, the performance of DiffNet and STAR-GCN degrades more quickly than that of AGNN. The main reason is that DiffNet and STAR-GCN are dependent on user-item interaction graph, and thus are sensitive to the number of cold start nodes which is proportional to the number of edges in the user-item  graph. In contrast, AGNN focuses on modeling the  attribute graph and is less affected by the limited number of interactions.


\section{Conclusion}
In this paper, we propose a novel model, namely AGNN for cold start recommendation. We first highlight the importance of exploiting the attribute graph rather than the interaction graph in addressing cold start problem in neural graph recommender systems. We then present an eVAE structure to infer  preference embedding from attribute distribution. Moreover, we address the key challenges in aggregating various information in a neighborhood by developing a gated-GNN structure which greatly improves the model capacity.
We conduct extensive experiments on two real world datasets. Results prove that our AGNN model outperforms the state-of-the-art methods for cold start recommendation. It  also achieves better or competitive  performance than these baselines in warm start scenario.

\subsubsection{Acknowledgments.}
This work has been supported in part by the NSFC Projects (61572376, 91646206).


\end{document}